# The *SVOM* gamma-ray burst mission

B. Cordier[*,1], J. Wei[2], J.-L. Atteia[4], S. Basa[5], A. Claret[1], F. Daigne[6], J. Deng[2],
Y. Dong[3], O. Godet[4], A. Goldwurm[1,7], D. Götz[1], X. Han[2], A. Klotz[4], C. Lachaud[7],
J. Osborne[9], Y. Qiu[2], S. Schanne[1], B. Wu[3], J. Wang[2], C Wu[2], L. Xin[2], B. Zhang[8],
S.-N. Zhang[3]

[1] *CEA Saclay, DSM/IRFU/service d'Astrophysique, 91191, Gif-sur-Yvette, France*
[2] *National Astronomical Observatories/CAS, 20A Datun Road, Beijing, 100012, China*
[3] *Institute of High Energy Physics/CAS, 19B YuquanLu, Beijing, 100049, China*
[4] *Institut de Recherche en Astrophysique et Planétologie (IRAP), UPS-OMP, Toulouse, France*
[5] *Laboratoire d'Astrophysique de Marseille, 38 rue Juliot-Curie, 13388 Marseille, France*
[6] *Institut d'Astrophysique de Paris, 98 bis boulevard Arago, 75014 Paris, France*
[7] *Astroparticule et Cosmologie (APC), 75013 Paris, France*
[8] *Department of Physics, University of Nevada, Las Vegas, NV 89154, USA*
[9] *Department of Physics and Astronomy, University of Leicester, Leicester, LE1 7RH, UK*

*E-mail:* `b.cordier@cea.fr`

We briefly present the science capabilities, the instruments, the operations, and the expected performance of the SVOM mission. SVOM (Space-based multiband astronomical Variable Objects Monitor) is a Chinese-French space mission dedicated to the study of Gamma-Ray Bursts (GRBs) in the next decade. The SVOM mission encompasses a satellite carrying four instruments to detect and localize the prompt GRB emission and measure the evolution of the afterglow in the visible band and in X-rays, a VHF communication system enabling the fast transmission of SVOM alerts to the ground, and a ground segment including a wide angle camera and two follow-up telescopes. The pointing strategy of the satellite has been optimized to favor the detection of GRBs located in the night hemisphere. This strategy enables the study of the optical emission in the first minutes after the GRB with robotic observatories and the early spectroscopy of the optical afterglow with large telescopes to measure the redshifts. The study of GRBs in the next decade will benefit from a number of large facilities in all wavelengths that will contribute to increase the scientific return of the mission. Finally, SVOM will operate in the era of the next generation of gravitational wave detectors, greatly contributing to searches for the electromagnetic counterparts of gravitational wave triggers at X-ray and gamma-ray energies.

[*]Speaker

## 1. Introducing SVOM

Gamma-ray bursts (GRBs) are powerful cosmic explosions associated with the end of life of some massive stars or with the merging of two compact stars. In both cases the end product of the explosion is a black hole or a young magnetar surrounded by a torus of matter that is quickly accreted onto the compact object (in seconds), releasing huge amounts of energy in two transient relativistic jets. When one of the jets is pointed to the Earth, we see a bright high energy transient followed by a quickly fading afterglow. A general introduction to these events can be found in [1,2, 3].

The study of GRBs sheds light on several key questions of modern astrophysics, like the physics at work in astrophysical relativistic jets, the end of life of massive stars and the birth of stellar mass black holes, the history of massive star formation over the ages, the history of the reionization of the universe and of the chemical enrichment of galaxies, etc. Some of these questions are directly connected with the study of GRBs, while other use them as lighthouses illuminating the remote regions of our universe, out to redshift $z \geq 10$ [4]. The detection of GRBs will thus remain of high priority in the coming years.

It is remarkable that all GRBs known to date have been detected from space with their prompt high energy emission, while all redshifts have been measured with large telescopes on the ground. This emphasizes the necessary synergy between space and ground in the study of these events and in their use as cosmological tools. Even if the direct detection of optical afterglows from the ground may become trivial in the future, with powerful telescopes like LSST, pursuing the many scientific objectives listed above requires the detection of the prompt high energy emission, which can only be done from space. This is the best way to measure the energy radiated by the GRBs and to get timely alerts for the measure of the redshift, which must happen within few hours of the burst when the optical afterglow is sufficiently bright.

The success of the *Swift* mission for catching GRBs and other transients illustrates the benefit of its unique combination of space agility, fast communication with the ground, multiwavelength observing capability, and long lifetime. These capabilities have permitted the detection of nearly 1000 GRBs and the measure of nearly 300 redshifts, have offered a new look at GRB progenitors and have led to several important discoveries, like the existence of GRBs at the epoch of the reionization of the universe.

The aim of SVOM (Space-based multiband astronomical Variable Objects Monitor) is to continue the exploration of the transient universe with a set of space-based multi-wavelength instruments, following the way opened by *Swift*. SVOM is a space mission developed in cooperation between the Chinese National Space Agency (CNSA), the Chinese Academy of Science (CAS) and the French Space Agency (CNES). The mission features a medium size satellite, a set of space and ground instruments designed to detect, locate and follow-up GRBs of all kinds, a anti-Sun pointing strategy allowing the immediate follow-up of SVOM GRBs with ground based telescopes, and a fast data transmission to the ground. The satellite carries two wide field high energy instruments: a coded-mask gamma-ray imager called ECLAIRs, and a gamma-ray spectrometer called GRM, and two narrow field telescopes that can measure the evolution of the afterglow after a slew of the satellite: an X-ray telescope called MXT and an optical telescope called VT. The ground segment includes additional instrumentation: a wide angle optical camera (GWAC) monitoring the field of view of ECLAIRs in real time during part of the orbit, and two 1-meter class robotic follow-up telescopes (the GFTs). SVOM has some unique features: an energy threshold of ECLAIRs at 4 keV enabling the detection of faint soft GRBs (e.g. XRFs and high-z GRBs); a good match in sensitivity between the X-ray and optical space telescopes which permits the detection of most GRB afterglows with both telescopes; and a set of optical instruments on the ground dedicated to the mission. The mission has recently been confirmed by the Chinese and French space agencies for a launch in 2021, and it has entered in an active phase of construction. This paper presents an overview of the scientific objectives of SVOM in the next section, and a brief description of the instrumental aspects of the mission in sections 3 and 4.The reader interested in more details can refer [5] and .[6].

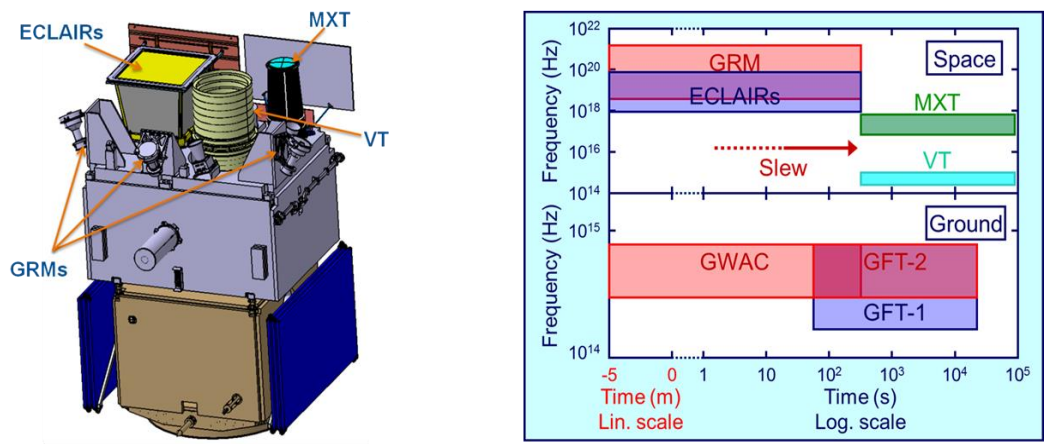

**Figure 1**: Schematic showing the SVOM spacecraft with its multi-wavelength space payload. It consists of two wide-field instruments: ECLAIRs & the Gamma-Ray Monitor (GRM) for the observation of the prompt emission and two narrow field instruments: the Micro channel X-ray Telescope (MXT) and the Visible Telescope (VT) for the observation of the afterglow emission. Right: Space and ground instruments join to enable a unique coverage in time and wavelength.

## 2.    Scientific objectives

The main science goal of SVOM is the study of cosmic transients detected in hard X-rays and in the optical. While the mission has been designed for the study of GRBs, it is also well suited for the study of other types of high-energy transients like tidal disruption events, active galactic nuclei, or galactic X-ray binaries and magnetars. For this type of sources, SVOM is both a "discovery machine", with wide field instruments that survey a significant fraction of the sky (ECLAIRs, GRM and GWAC), and a "follow-up machine", with fast pointing telescopes in space and on the ground (MXT, VT, and GFTs) that provide a multi-wavelength follow-up of remarkable sources, with good sensitivity and a high duty cycle. The follow-up can be triggered by the satellite itself or from the ground, upon reception of a request for target of opportunity observations (ToO). The science goals of SVOM are presented in the document "SVOM: a space mission to probe fundamental physics and to explore the young universe".

In this section we concentrate on the objectives of SVOM for GRBs, and we refer the reader to the aforementioned document for a description of the science goals concerning other sources. We have evaluated a GRB detection rate of 70-80 GRBs/yr for ECLAIRs and $\geq 90$

GRBs/yr for GRM. One essential goal of SVOM is to get GRBs with a redshift. The redshift is required to measure the energetics of the burst and the epoch at which the GRB occurred in the history of the universe. Four elements in the design of the mission concur to facilitate the measure of the redshift for SVOM GRBs: a near anti-solar pointing ensuring that SVOM GRBs can be quickly observed with ground based telescopes, a good sensitivity of the on-board optical telescope permitting the rapid identification of high-z candidates (which are not detected at visible wavelengths), NIR follow-up on the ground to look for the afterglows of dark GRBs, and agreements with the community to promote the optical spectroscopy of SVOM GRBs with large telescopes. With this strategy we expect to measure the redshift of more than 50% of SVOM GRBs, constructing a sample that could be more representative of the true GRB population than the Swift sample. One drawback of this pointing strategy is that it avoids the galactic plane. Galactic sources can nevertheless be observed with target of opportunity pointings. In the following, we discuss a few selected topics about GRBs for which SVOM may bring decisive progress.

### 2.1 GRB Progenitors

In order to get a better understanding of the GRB phenomenon SVOM was designed to detect all kinds of GRBs, and to provide extensive multi-wavelength observations of the prompt GRB and its afterglow. In addition to classical long GRBs that will be detected by the two wide field instruments (ECLAIRs and GRM), SVOM will routinely detect short bursts with its gamma-ray monitor and soft GRBs (XRFs or highly redshifted GRBs) with ECLAIRs. The ability to detect soft GRBs with a spectral energy distribution (SED) peaking below 20-30 keV will favor the detection of X-ray flashes. As shown by BeppoSAX and HETE-2 [7, 9], these faint GRBs can only be detected if they are close enough and if their spectral energy distribution peaks at low energies (because faint GRBs with a SED peaking at high energies radiate too few photons to be detected). The detection of X-Ray Flashes in the local universe ($z \leq 0.1$) will permit detailed studies of their associated supernovae (as was the case for XRF 020903 [10]), providing crucial clues to understand the broader context of the SN-GRB connection.

### 2.2 GRB Physics

The instrument suite of SVOM will provide good multi-wavelength coverage of GRBs. For those occurring in the field of view of GWAC, the prompt emission will be measured from 1 eV to 5 MeV, with GWAC, ECLAIRs, and GRM. The prompt optical emission may also be observed by one GFT for GRBs lasting longer than 40 seconds. GRB afterglows will be observed with the two narrow field instruments on-board the satellite and with the ground follow-up telescopes on Earth. For some GRBs, SVOM will provide a very complete view of the phenomenon and its evolution, hopefully bringing new insight into the complex physics at work in these events. A lesson learned from Swift is that a few well observed GRBs may crucially improve our understanding of GRB physics, as was the case for GRB 130427, a bright nearby burst detected by Swift, Fermi and various optical and radio telescopes on the ground [11,13].

The physical processes at work within the jet remain not well understood even after the observation of hundreds of GRBs. Comprehensive discussions of the theoretical challenges connected with the understanding of the prompt GRB emission can be found [14] and [15]. The authors show that several crucial questions connected to the physics of the ultra-relativistic jet and its interaction with the surrounding medium remain unanswered, like the nature and content of the jet (is the energy stored in the baryons or in magnetic fields?), the mechanisms of particle acceleration, the microphysics and the dominant radiation processes, the importance of the reverse shock, the role of pairs, etc. Performing multi-wavelength observations during the prompt emission and the early afterglow, SVOM will provide key observations to understand the physics of relativistic jets. GRBs detected with SVOM will also benefit from contemporaneous or follow-up observations with a novel generation of powerful instruments, like CTA (the Cerenkov Telescope Array) for very high energy photons, LSST (Large Synoptic Survey Telescope) for optical transients associated with on-axis and off-axis GRBs, and the precursors of SKA (the Square Kilometer Array) in radio.

### 2.3 Cosmology

GRBs are like "fireworks" in the distant universe. Their extreme luminosity permits their detection in hard X-rays up to very high redshifts ($z > 10$) and the spectroscopy of the optical afterglow provides the redshift of the burst and a tomographic vision of the line of sight to the burst. With Swift and the measure of about 300 redshifts, GRBs are providing new diagnostics of the distant universe. When the signal to noise ratio (SNR) of the optical spectrum of the afterglow is sufficient, we get detailed information on the circumburst medium, on the gas and dust in the host galaxy, on the intergalactic medium and the intervening systems. With a smaller SNR, the measure of the redshift allows locating the time of the explosion in the history of the universe and reconstructing the history of the GRB formation rate, which traces the formation rate of massive stars.

There is of course a special interest in very distant GRBs ($z > 5$), which provide a unique view on the young universe, especially since they occur in galaxies which are undetectable with other methods of observation. One exciting challenge of GRB missions is the detection of GRBs resulting from the explosion of population III stars (the first generation of stars formed with pristine gas containing no metals). Such events are expected to be rare, to occur at high redshifts, to have no detectable hosts (except in absorption in the spectrum of their optical afterglow) and afterglows that are only detectable in the near infrared. They could be similar to some very long GRBs found at lower redshift [16, 17]. We expect to detect about 5 GRBs/yr at redshift $z > 5$ with ECLAIRs, but they will be useful only if we can measure their redshift. One difficulty is that the optical afterglows of GRBs fade very quickly, and after a few hours they are often too faint to permit measuring the redshift of the burst. In order to quickly identify high-z candidates that deserve deep spectroscopy in the NIR, we rely on the sensitivity of VT and on fast NIR follow-up telescopes on the ground. Dark GRBs, whose afterglows are not detected in the VT, are good candidates for high-z bursts, but they can also be extinct by dust in the vicinity of the source. The nature of these events (distant or extinct GRB) will be confirmed quickly with fast visible/NIR photometry from the ground, allowing the most appropriate spectroscopic follow-up. In some cases, GRBs detected by SVOM could trigger follow-up observations by the JWST (James Webb Space Telescope) for the study of the hosts.

### 2.4 Gravitationnal waves sources

SVOM will also contribute to clarify the origin of short GRBs, especially with the possibility to search for GRBs in coincidence with the signals detected by advanced gravitational waves (GW) detectors. The favorite scenario for the production of short GRBs is the coalescence of two compact objects (two neutron stars or a neutron star plus a black hole), which predicts that short GRBs are accompanied by powerful bursts of gravitational waves. One complication of these searches is that GRBs are much more beamed than gravitational waves. Considering that we detect 1 short GRB out of ~50 mergers [28; 29], and a moderate enhancement of the gravitational wave emission along the jet, we expect that about 10% of the mergers detected by advanced detectors of gravitational waves could be associated with a GRB. Assuming a rate of binary mergers of ~50/yr within the horizon of GW detectors (~400 Mpc), we expect to detect with ECLAIRs ~3 events coincident with GW triggers, and ~9 with GRM, in 5 years of operation. SVOM will also have the capability to point its narrow field instruments towards candidate sources of GWs. We evaluate that 15 events approximately can be followed quickly (6 hours) with SVOM narrow field instruments in 5 years of operation.

### 3. SVOM - Space based instruments

In this section and the following, we give a brief description of the instruments of SVOM. In order to be both a discovery machine and a follow-up machine, SVOM features two types of instruments: space based instruments and ground based instruments. We briefly describe these instruments below. A more complete description of the French payload can be found in [18].

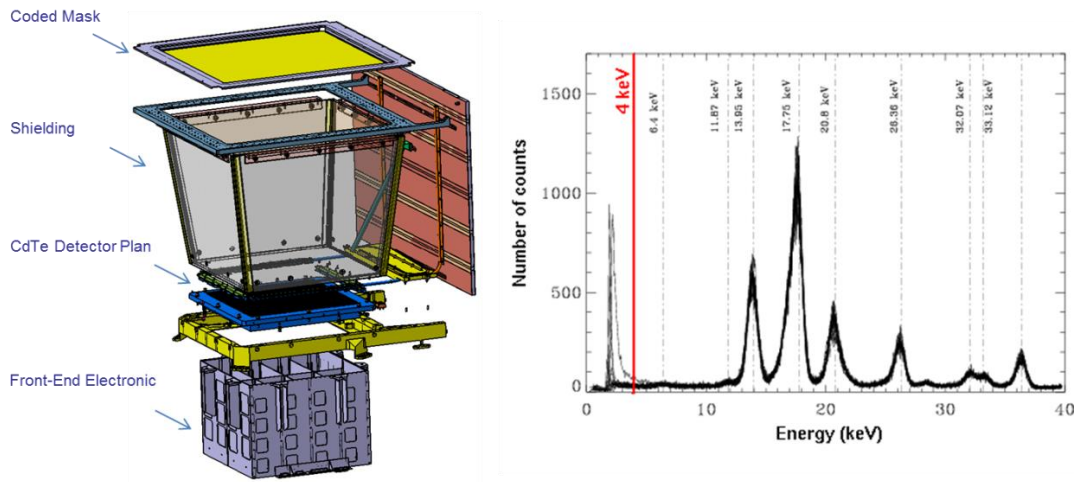

**Figure 2**: Schematics showing the different sub-systems of ECLAIRs except the data processing unit in charge of the GRB detection and localisation. Right: Laboratory spectral measurements performed on a detector module prototype (a 32 CdTe pixel matrice) with radioactive sources ($^{241}$Am). The red vertical line corresponds to the expected low energy threshold of the ECLAIRs camera [6].

### 3.1 ECLAIRs the hard X-ray coded mask imaging camera

ECLAIRs is the instrument onboard the satellite that will detect and locate the GRBs. ECLAIRs is made of four parts: a pixelated detection plane (1024 cm$^2$) with its readout electronics, a coded mask, a shield defining a field of view of 2 steradians (89° x 89°), and a processing unit in charge of detecting and locating transient sources. The detection plane is made of 200 modules of 32 CdTe detectors each, for a total of 6400 detectors of size 4x4x1 mm. Each module is read by a customized ASIC connected to an electronics that encodes the position, the time and the energy of each photon. One of the requirements of ECLAIRs is to reach an energy threshold of 4 keV, in order to study soft GRBs like X-Ray Flashes and highly redshifted GRBs. As shown in figure 2, the first modules satisfy the low energy threshold requirement [19-21].

The coded mask is a square of side 54 cm located at a distance of 46 cm from the detection plane; it has an opening fraction of 40% and provides a localization accuracy of several arcminutes (~14' for a source at the limit of detection). The instrument features a count rate trigger and an image trigger, like Swift. These triggers are computed, from the photon data, in several energy bands and on timescales ranging from 10 ms to several minutes. Our simulations show that ECLAIRs will detect 70-80 GRBs/yr. For more details on the trigger process, see Antier et al. (these proceedings) and for a complete description of ECLAIRs see Schanne et al. (these proceedings).

### 3.2 GRM: the Gamma-ray Monitor

GRM consists of a set of three detection modules. Each of them is made of a scintillating crystal (sodium iodide), a photomultiplier and its readout electronics. Each detector has a surface area of 200 cm$^2$ and a thickness of 1.5 cm. One piece of plastic scintillator in front of NaI(Tl) is used to distinguish low energy electrons from normal X-rays. The three modules are pointed at different directions to form a total field of view of 2 steradians, within which rough localization of transient sources can be achieved onboard. The energy range of the GRM is 15-5000 keV, extending the energy range of ECLAIRs towards high energies to measure Epeak for a large fraction of SVOM GRBs. We expect that GRM will detect > 90 GRBs/yr. GRM will have a good sensitivity to short/hard GRBs, like the GBM of Fermi. GRM can generate onboard GRM-only triggers, taking use of only GRM detectors. Such triggers with localization information will be transferred to ECLAIRs for trigger enhancement on the short GRBs, and to ground facilities (e.g. GWAC, GW experiments) for joint observations.

A calibration detector containing one radioactive $^{241}$Am isotope is installed on the edge of each detection module, for the purpose of gain monitoring and energy calibration. In addition, a particle monitor auxiliary to GRM can generate South Atlantic Anomaly alerts and help protecting the detection modules.

### 3.3 MXT: the Microchannel X-ray Telescope

MXT is a light and compact focusing X-ray telescope designed to observe and measure the properties of the GRB X-ray afterglow after a slew of the satellite. The telescope will implement a novel technique to focus X-rays, based on micropore optics arranged in a lobster eye geometry. The use of micropore optics instead of classical X-ray electro-formed mirrors permits a significant reduction of the size and weight of the telescope, fitting on a medium size satellite like SVOM. The optics has a diameter of 24 cm and a focal length of 1 meter [22].

MXT will make use of the radiation hard pn-CCD detector developed for the DUO mission and adopted in a larger version for the eROSITA mission [23;24]. The MXT detector is made of 256 x 256 Si pixels of 75 µm side and has an expected energy resolution of 75 eV at 1 keV.

MXT will be operated in the energy range of 0.2-10 keV, will have an effective area of 45 cm$^2$ at 1 keV, and a field of view of 64x64 arc minutes. Despite the smaller effective area with respect to XRT on board *Swift* at 1 keV (about 120 cm$^2$) MXT will be able to fully characterize the GRBs light curves (as shown in Fig.3). In fact with a sensitivity of 7 10$^{-13}$ erg cm$^{-2}$ s$^{-1}$ in 10$^4$ s, MXT will detect the afterglows of more than 90% of SVOM GRBs. Indeed its sensitivity is well adapted to early GRB afterglow observations, and its PSF of about 4 arcmin will allow to significantly reduce the ECLAIRs error boxes. Simulations based on the *Swift*-XRT database show that the localization accuracy for the MXT is ~13" for 50% of the bursts, 10 minutes after the trigger (statistical uncertainty only). For more details on the MXT instrument, see Götz et al. (these proceedings).

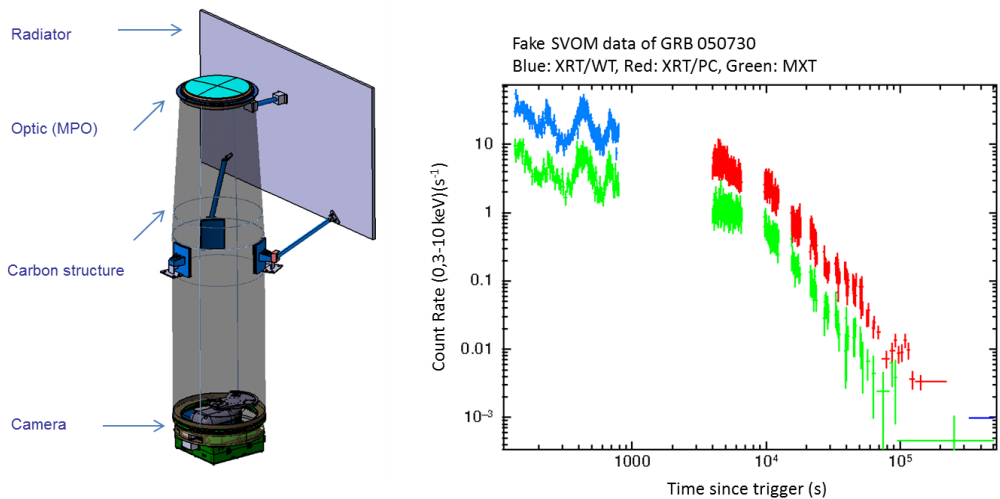

**Figure 3**: Schematics showing the different sub-systems of the MXT. Right: XRT measured (blue and red points) versus MXT simulated (green points) light curve of GRB 050730, a GRB with a median flux in the Swift/XRT afterglow database.

### 3.4   VT: the visible Telescope

The Visible Telescope (VT) is a dedicated optical follow-up telescope on board the SVOM satellite. Its main purpose is to detect and observe the optical afterglows of gamma-ray bursts localized by ECLAIRs. It is a Ritchey-Chretien telescope with a diameter of 40 cm and an f-ratio of 9. Its limiting magnitude is about 22.5 (Mv) for an integration time of 300 seconds

VT is designed to maximize the detection efficiency of GRB's optical afterglows. Instead of a filter wheel, a dichroic beam splitter is used to divide the light into two channels, in which the GRB afterglow can be observed simultaneously. Their wavelength ranges are from 0.4µm to 0.65µm (blue channel) and from 0.65µm to 1µm (red channel). Each channel is equipped with a 2k x 2k CCD detector. While the CCD for the blue channel is a normal thinned back-illuminated one, a deep-depleted one is adopted for the red channel to obtain high sensitivity at long wavelength. The Quantum Efficiency (QE) of the red-channel CCD at 0.9µm is over 50%, which enables VT to have the capability of detecting GRBs with the redshift larger than 6.5. The field of view of VT is about 26'x 26', which can cover the error box of Eclairs in most cases. Both CCDs have a pixel size of 13.5µm x 13.5µm, corresponding to spatial resolutions of 0.77 arc-second. This ensures the GRB positioning accuracy to be greatly improved by VT from several arc-minutes (Eclairs) and tens arc-seconds (MXT) to a level of sub-arcsecond.

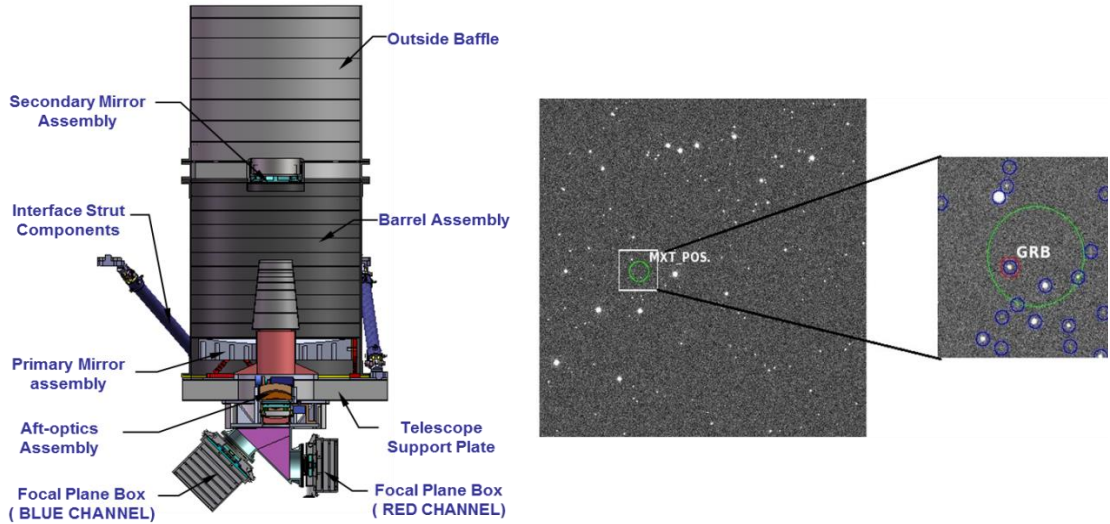

**Figure 4**: Schematic showing the different sub-systems of the VT. Right: a simulated image of theVT and its associated finding chart.

In order to promptly provide the GRB alerts with the sub-arcsecond accuracy, VT will do some data processing on board. After a GRB has been localized by the coaligned MXT, lists of sources are extracted from the VT sub-images whose centers and sizes are determined by the GRB positions and the corresponding error boxes provided by MXT. The lists are immediately downlinked to the ground through the VHF network. Then, the ground software will make finding charts with these lists (see Fig.4) and search the optical counterparts of the GRB by comparing the lists with the existing catalogues. If a counterpart is identified, an alert will be then produced and distributed to world-wide astronomical community, which is useful for triggering large ground-based telescopes to measure the redshifts of the GRBs by spectroscopy.

VT is expected to do a good job on detecting high-redshift GRBs. The confirmed high-redshift GRBs are rare in the *Swift* era, in contrast to a theoretical prediction of a fraction of more than 10%. This is probably due to the fact that for most *Swift* GRBs the early-time optical imaging follow-up is not deep enough for a quick identification and some faint GRBs cannot be spectroscopically observed in time by the large ground-based telescopes. This passive situation will be significantly improved by SVOM, due to the high sensitivity of VT, in particular at long wavelength, and the prompt optical-counterpart alerts. Additionally, the anti-solar pointing strategy of SVOM allows GRBs to be spectroscopically observed by large ground-based telescopes at the early time of the bursts. Consequently, more high-redshift GRBs are expected to be identified in the SVOM era.

VT is also used to support the platform to achieve the required high pointing stability. A Fine Guidance Sensor (FGS) is mounted on the VT focal plane to measure relative image motions. Its images are processed in real time by a specialized data processing unit to get the centroid positions of several stars brighter than the magnitude of 15 (Mv). The results are sent at a frequency of 1Hz to the platform to improve the pointing stability, which enables VT to have a good performance in a long exposure time.

## 4. Instruments on the ground

The ground follow-up instruments constitute an important part of the mission. Three instruments are developed for the follow-up of SVOM GRBs: a wide angle camera that surveys a significant fraction of the sky for transients, and two robotic telescopes. In addition to these dedicated instruments, the SVOM collaboration will seek agreements with various existing telescopes or networks willing to contribute to the follow-up of SVOM GRBs.

### 4.1 GWAC: the Ground Wide Angle Camera

GWAC provides a unique way to survey a large field of view for optical transients. The instrument will monitor 63% of ECLAIRs field of view, looking for optical transients occurring before, during and after GRBs. GWAC will also have its own trigger system, providing alerts to the world. GWAC is a complex system: the heart of the system is a set of 36 wide angle cameras with a diameter of 18 cm and a focal length of 22 cm, together these cameras cover a field of view of 5000 sq. deg. They use 4k x 4k CCD detectors, sensitive in the range of wavelength 500-800 nm. These cameras reach a limiting magnitude V=16 (5 $\sigma$) in a typical 10 second exposure. This set of cameras is completed by two 60 cm robotic telescopes. equipped with EMCCD cameras. These telescopes will provide multicolor photometry of the transients discovered by GWAC with a temporal resolution $\leq$ 1 second.

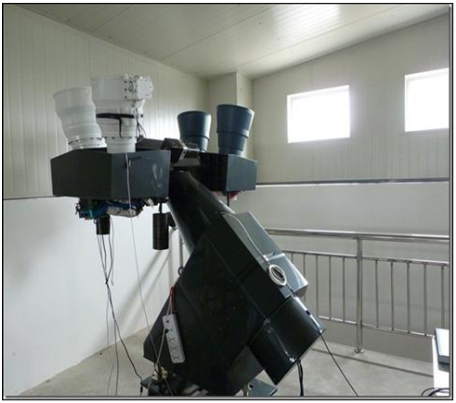
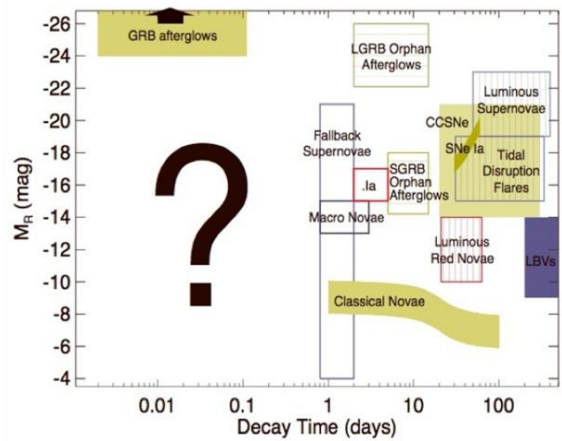

**Figure 5**: First prototype of one GWAC module. The final system will be composed by 9 modules. Right: the figure shows the dicovery space of an instrument dedicated to short time-scale optical transients [30].

### 4.2 GFTs: the Ground Follow-up Telescopes

The ground follow-up telescopes have two main goals. Firstly, they measure the photometric evolution of the optical afterglow in the first minutes after the trigger in a broad range of visible and NIR wavelengths, with a temporal resolution of few seconds. Secondly, when an afterglow is detected, they provide its position with arcsecond precision within 5 minutes of the trigger. Some essential features of the GFTs are their field of view, their size, and their sensitivity in the near infrared. The field of view (~30 arcminutes) enables observing quickly the entire error boxes of ECLAIRs. The size, typically 1 meter, allows the detection of all visible (i.e. non-dark) afterglows at the condition to arrive within few minutes after the trigger [25; 26]. Finally, the near infrared sensitivity permits the detection of high-z GRBs and GRBs extinct by dust, whose afterglow are obscured in the visible domain [27]. GFTs are especially useful for the study of the early afterglow during the slew of the satellite, and for the rapid identification of the optical afterglow in various cases: when SVOM cannot slew to the burst or when the slew is delayed due to pointing constraints, and when the optical afterglow is only visible in the NIR.

One telescope is located in China at Xinglong Observatory and one telescope will be located in Mexico at San Pedro Mártir.

## 5. The system

In order to facilitate measuring the redshifts of GRBs detected with ECLAIRs, the instruments of SVOM will be pointed close to the anti-solar direction. Most of the year the optical axis of the SVOM instruments will be pointed at about 45° from the anti-solar direction. This pointing is interleaved with avoidance periods during which the satellite passes away from the Sco-X1 source and the galactic plane. This strategy ensures that SVOM GRBs will be in the night hemisphere and quickly observable from the ground by large telescopes. More details on the SVOM pointing strategy, can be found in [31].

As soon as a GRB will be located, its coordinates and its main characteristics will be sent to the ground within seconds with a VHF antenna. The VHF signal will be received by one of the ~ 40 ground stations distributed around the Earth below the orbit. The data will then be relayed to the Operation Center, which will send SVOM alerts to the internet via the GCN and VOEvent networks (http://gcn.gsfc.nasa.gov/), and to the ground instruments GWAC and the GFTs. SVOM can also perform target of opportunity observations (with MXT and VT for instance), with a delay of few hours, which depends on the availability of uplink communication with the satellite.

## 6. Conclusion

SVOM, like Swift, will be a highly versatile astronomy satellite, with built-in multi-wavelength capabilities, autonomous repointing and dedicated ground follow-up. We look forward to a broad science return of SVOM thanks to its unique instrumental combination of three wide-field instruments: ECLAIRs, GRM, and GWAC, and three narrow-field instruments: MXT, VT, and GFTs. Beyond the GRB studies emphasized here, SVOM will bring new observations about all types of high energy transients, in particular those of extragalactic origin (TDEs, AGNs, etc.). The possibility to detect and localize short GRBs associated with gravitational wave events appears especially exciting, even if it is challenging.

## Acknowledgments

We acknowledge the contribution of the SVOM project Group at SECM Shanghai and CNES Toulouse. Y. Qiu and J. Wang are supported by the National Natural Science Foundation of China (Nos. U1231115 and 11473036).